\documentclass[%
 reprint,
 amsmath,amssymb,
 aps,
]{revtex4-2}

\usepackage{graphicx}
\usepackage{multirow}
\usepackage{amsmath} 
\usepackage{dcolumn}
\usepackage{bm}
\usepackage{xcolor}

\begin{document}

\preprint{APS/123-QED}

\title{Thermal Model for Time-Domain Thermoreflectance \\
Experiments in a Laser Flash Geometry}

\author{Wanyue Peng}
\author{Richard B. Wilson}%
 \email{rwilson@ucr.edu}
\affiliation{%
 Mechanical Engineering and Materials Science, University of California Riverside\\
}%





\begin{abstract}
\noindent
Time-domain thermoreflectance (TDTR) is a well-established pump/probe method for measuring thermal conductivity and interface conductance of multilayers.
Interpreting signals in a TDTR experiment requires a thermal model. 
\textcolor{black}{
In standard front/front TDTR experiments, both pump and probe beams typically irradiate the surface of a multilayer. As a result, existing thermal models for interpreting thermoreflectance experiments assume the pump and probe beams both interact with the surface layer.}
Here, we present a frequency-domain solution to the heat-diffusion equation of a multilayer in response to nonhomogenous laser heating.  
This model allows analysis of experiments where the pump and probe beams irradiate opposite sides of a multilayer. 
We call such a geometry a front/back experiment to differentiate such experiments from standard TDTR experiments. 
As an example, we consider a 60nm amorphous Si film. We consider how signals differ in a front/front vs. front/back geometry and compare thermal model predictions to experimental data. 
\end{abstract}

\maketitle

\section{Introduction}
\noindent
Optical techniques for studying thermal transport properties require an accurate thermal model for heat-flow in a layered geometry \cite{cahill2004analysis,jiang2018tutorial,yang2016modeling,dilhaire2011heterodyne,yuan2019nanosecond}. Examples of such optical methods include time-domain thermoreflectance \cite{cahill2004analysis,feser2012probing,feser2014pump}, frequency-domain thermoreflectance \cite{regner2013broadband,rodin2017simultaneous,schmidt2009frequency}, thermoreflectance microscopy \cite{ziabari2018full,farzaneh2009ccd,maize2014high, dallas2018thermal}, and steady-state thermoreflectance \cite{braun2019steady}.
Raman thermometry \cite{beechem2015invited,kargar2020phonon} also requires an accurate prediction of the temperature response of a multilayer to optical heating. 
\noindent
\\
Thermal models used to interpret thermoreflectance experiments typically assume that heat is absorbed, and the temperature is measured at the sample’s surface \cite{cahill2004analysis,feser2012probing,feser2014pump}. Laser heating is treated as a heat-current boundary condition at the multilayer’s surface. Thermoreflectance signals are assumed to be proportional to the temperature at the surface of the top layer. While neither assumption is precisely correct, the resulting error from these assumptions is usually negligible in a normal front/front thermoreflectance experiment. 
By front/front, we mean both the pump and probe laser beams are incident at the top surface of the multilayer. 
\textcolor{black}{The assumption that heating occurs at a single plane will cause an error if the length-scale over which heat is absorbed has enough thermal resistance to affect the signal at the time scales of interest. This can happen if the metal’s thermal conductivity is low \cite{wilson2015thermal} or in experiments with no metal film transducer \cite{yang2016modeling,braun2019steady,zenji2020ultimate}.}

\noindent
\\
A number of recent papers describe analytic solutions to the heat-diffusion equation in a layered geometry that don’t assume heat is deposited at the surface. \textcolor{black}{
Olson et al. \cite{olson2019spatially} describes how the optical penetration depth affects thermoreflectance signals.} Wilson et al. \cite{wilson2013two} and Regner et al. \cite{regner2015interpretation} describes a two-temperature model where heat can be deposited at a buried interface. Similarly, Braun et al. \cite{braun2017upper} describes a single temperature bidirectional model where heat is deposited, and temperature is measured, at a buried interface. A number of thermoreflectance studies account for the effects of non-homogeneous heating by linearly combining solutions from such bidirectional models \cite{kimling2015spin,wilson2015thermal}. Yang et al. \cite{yang2016modeling} modified the standard thermoreflectance thermal model to allow for an exponentially decaying volumetric heat source in the top-most layer. 
\textcolor{black}{
A limitation common to many of these thermal models is the assumption that heating and temperature measurement occur in the same layer. Still missing from the literature is a heat-equation solution that allows heating of any functional form across the entire multilayer, and that does not assume temperature is measured in the same layer where heating occurs.}

\noindent 
\\
Here, we generalize the analytic solution for heat flow in a layered structure \cite{cahill2004analysis} to allow for non-homogeneous optical heating across the layered structure.
Our primary motivation is the analysis of thermoreflectance data collected in a laser flash geometry, i.e., a front/back geometry.  In a front/back geometry, the pump and probe beams impinge on opposite sides of the multilayer structure of interest \cite{gomez2020high,choi2014indirect}.
\textcolor{black}{
Therefore, a thermal model that treats the position of the heater and thermometer as input parameters is necessary for the analysis of TDTR experiments in a front/back geometry.}
By allowing the heating as a function of depth to be an arbitrary function, our model allows analysis of such front/back experiments.
\textcolor{black}{
Additionally, the model accomplishes a secondary goal. The model allows for accurate modeling of the interaction length scales of the laser with the sample.} \\
\noindent \\
The outline of the paper is as follows.  First, we derive a frequency-domain solution to the heat diffusion equation in a layered geometry in response to non-homogeneous heating vs. depth.  Next, as a test case, we compare experimental data to our thermal model predictions for the frequency-domain and time-domain thermoreflectance measurements of \textcolor{black}{ a 60 nm a-Si layer sample}. Finally, we evaluate how signals in the front/back time-domain-thermoreflectance experiment depend on the thermal properties of the stack.\begin{figure*}[!ht]
  \centering
    \includegraphics[width=0.9\textwidth]{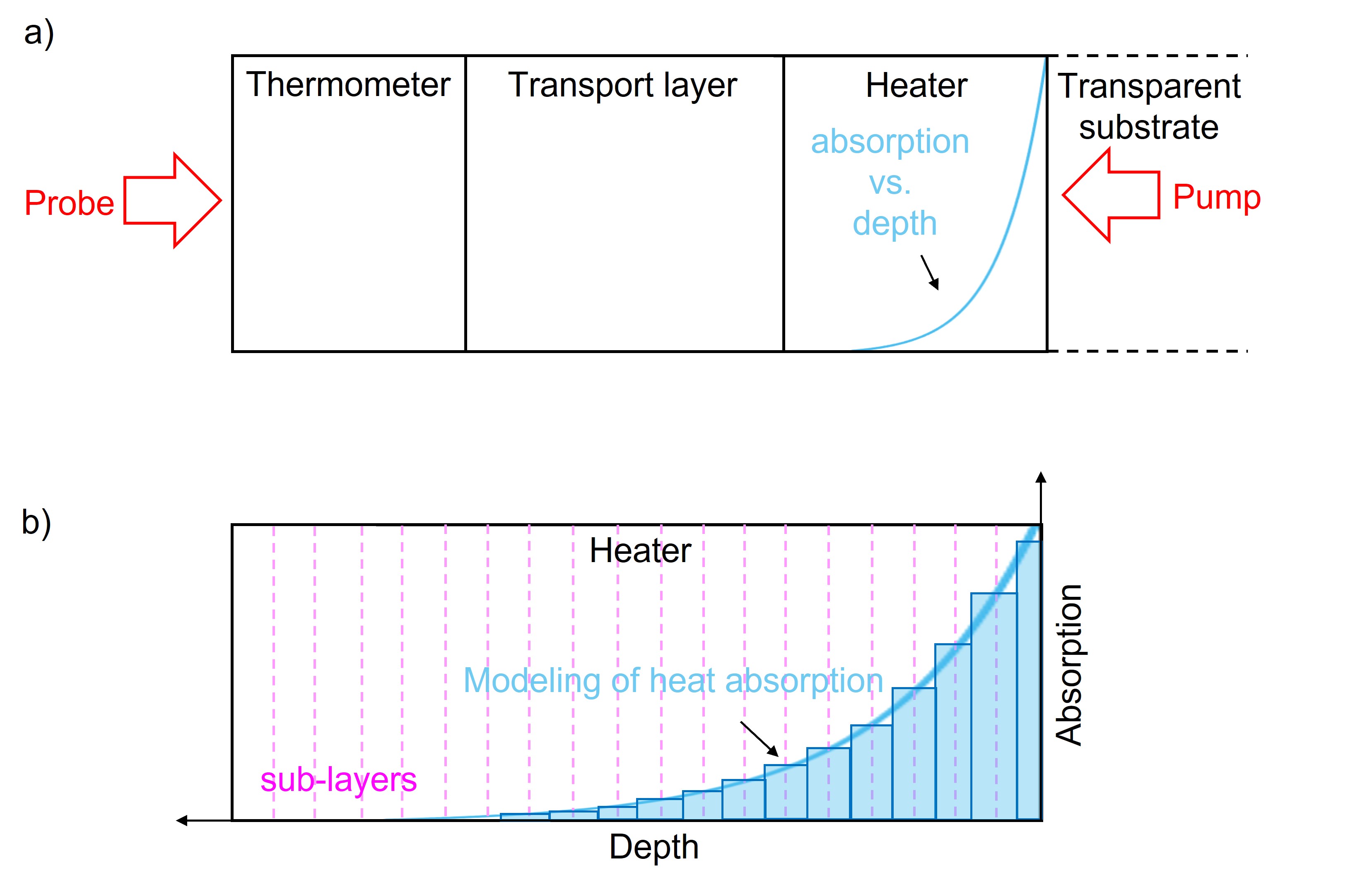}
    \caption{a) Schematic of pump/probe experiment in a laser flash geometry. In this geometry, the pump beam impinges through the sapphire and is absorbed by a buried heater layer. The blue curve represents the pump beam absorption as a function of depth.   b) Sketch of how we use closely spaced planar heat sources to approximate volumetric heating in our model. Layers that absorb pump energy are sub-divided into multiple layers in our thermal model. The heat-current $Q_n$  at each interface (pink dashed line) is determined by the area of the blue rectangles.}
     \label{schematic}
\end{figure*}
\section{Thermal model} \label{model}
\noindent
In response to harmonic heating at a frequency, $\omega$, heat flow in the $n^{th}$ layer of a multilayered structure is governed by the heat equation:
\noindent
\\
\begin{equation}
\frac{\partial^2 T_n}{\partial z^2} + \frac{\eta_n}{r}\frac{\partial }{\partial r}\left ( r\frac{\partial T_n}{\partial r} \right )= \frac{i \omega}{\alpha_n}T_n.
\label{heat_equation}
\end{equation}
 Eq. (\ref{heat_equation}) is written in cylindrical coordinates with an assumption of radial symmetry. Depth is described by $z$, and $r$ describes the radial distance from the center of the laser beam. $T_n$, and $\alpha_n$ are the temperature and thermal diffusivity of the $n^{th}$ layer. $\eta_n$ is the thermal anisotropy and is defined as the ratio of the radial thermal conductivity $\Lambda_r$ of layer $n$ and through-plane thermal conductivity of layer $n$, $\Lambda_n$. 

\noindent
\\
Taking the Hankel transform of Eq. (\ref{heat_equation}) turns it into an ordinary-differential equation whose solution is:
\begin{equation}
T_n(z) = B_{n}^{+}e^{u_nz} + B_{n}^{-}e^{-u_nz},
\label{Tn}
\end{equation}
where $u_n = \left [4\pi k^2\left ( \eta \right )+ \left (\frac{i\omega}{\alpha_n}\right )^{1/2} \right ]$. Here, $k$ is a spatial frequency coordinate that arises due to the Hankel transform of Eq. (\ref{heat_equation}).

\noindent
\\
According to Fourier's law, the heat-current at an arbitrary depth in layer $n$ is:
\begin{equation}
J_n(z) = - \gamma_n(B_{n}^{+}e^{u_nz} - B_{n}^{-}e^{-u_nz}),
\end{equation}
where $\gamma_n = \Lambda_nu_n$, and $\Lambda_n$ is the through-plane thermal conductivity of the corresponding layer.

\noindent
\\
Solving for the unknown temperature amplitudes $B_{n}^{+}$ and $B_{n}^{-}$ in all $N$ layers requires applying $2N$ boundary conditions. At each interface between two layers, we assume the temperature is continuous:
\begin{equation}
T_{n-1}(L_{n-1}) = T_{n}(0),
\label{boundary1}
\end{equation}
where $L_n$ is the thickness of the corresponding layer. 

\noindent
\\
Next, we assume some fraction of the laser intensity, $Q_n$, is deposited at each interface. With this assumption, energy conservation and Fourier's law implies:
\noindent
\begin{equation}
Q_{n}=-J_{n-1}(L_{n-1})+ J_n(0).
\label{boundary2}
\end{equation}

\noindent
\\
Here, $Q_n$ is a heat-current due to laser absorption in proximity to the interface at the top of layer $n$. We describe how the value of $Q_n$ is determined below.  

\noindent
\\
Applying the boundary conditions given by Eq. (\ref{boundary1}) and (\ref{boundary2}) allows us to relate the unknown temperature amplitudes in Eq. (\ref{Tn}) in two adjacent layers:
\begin{equation}
\begin{bmatrix}
B_{n-1}^{+}\\ 
B_{n-1}^{-}
\end{bmatrix} = \begin{bmatrix}
m
\end{bmatrix}_n \begin{bmatrix}
B_{n}^{+}\\ 
B_{n}^{-}
\end{bmatrix} + \begin{bmatrix}
S_{n}^{1}\\ 
S_{n}^{2}
\end{bmatrix},
\label{iteration}
\end{equation}

\noindent
\\
where
\begin{equation*}
\begin{bmatrix}
m
\end{bmatrix}_n= \frac{1}{2\gamma_n}\begin{bmatrix}
 e^{-u_nL_n}&0 \\ 
 0& e^{u_nL_n}
\end{bmatrix} \begin{bmatrix}
\gamma_n + \gamma_{n+1} & \gamma_n - \gamma_{n+1} \\ 
 \gamma_n - \gamma_{n+1} & \gamma_n + \gamma_{n+1}
\end{bmatrix}
\end{equation*}

\noindent
\\
and
\begin{equation*}
    \begin{bmatrix}
S_n^1\\ 
S_n^2
\end{bmatrix} = \frac{Q_n}{2\gamma_n}\begin{bmatrix}
e^{-u_nL_n} \\
-e^{u_nL_n}
\end{bmatrix}.
\end{equation*}

\noindent
\\
Compared with the solution for
front/front TDTR given by Cahill (2004) \cite{cahill2004analysis}, $\begin{bmatrix}
S_n^1\\ 
S_n^2
\end{bmatrix}$ is an additional term that accounts for the deposition of heat at interfaces throughout the multilayer. One of the goals for our thermal model is to accommodate arbitrary non-homogenous heating. We approximate volumetric heating vs. depth with a series of closely spaced planar heat sources. We sub-divide the layers where absorption takes place into a number of smaller layers.
Then, we set the heat-current $Q_n$ at each interface in the sub-divided absorbing layer based on the amount of volumetric heating nearby the interface's location. Absorption vs. depth of the pump beam is calculated using a multilayer optical calculation \cite{hecht1998optics}. A schematic illustration of how we sub-divide the layer or layers that absorb pump light and how we set $Q_n$ at the internal interfaces of those layers is shown in FIG. \ref{schematic}b. (In FIG. \ref{schematic}b, for simplicity, we illustrate a simple numerical integration scheme. Computational efficiency can be improved over what is sketched by subdividing the lasers based on more accurate numerical integration techniques, e.g., Gauss quadrature.)
\\
\noindent \\
An alternative way to accommodate non-homogenous heating would be to include volumetric heating terms directly in the heat-equation, Eq. (\ref{heat_equation}). This leads to additional particular solution terms in Eq. (2). Then, unknown temperature amplitudes can be determined by applying boundary conditions in a manner very similar to how we outline below. \textcolor{black}{This alternative approach leads to a somewhat more elegant, fully analytic final result. However, we found this fully analytic approach to be cumbersome to numerically evaluate across the wide spectrum of frequencies needed for TDTR analysis. 
Alternatively, the implementation of Eq. (6) is straightforward.
The approach illustrated in FIG. 1 can be viewed as a numerical approximation for volumetric heating.}\\
\noindent \\
The unknown temperature amplitudes of the surface layer, $B_1^+$ and $B_1^-$, are determined by iteratively applying Eq. (\ref{iteration}).  We start at the bottom layer farthest from the surface, i.e., $n=N$.  We assume the bottom layer is semi-infinite, therefore $B_N^+=0$. Then,

\begin{equation}
    \begin{bmatrix}
    B_1^+\\
    B_1^-
    \end{bmatrix}
    = [M]
    \begin{bmatrix}
    0\\
    B_N^-
    \end{bmatrix}
    + \begin{bmatrix}
    \delta_+\\
    \delta_-
    \end{bmatrix}.
\label{halfsolved}
\end{equation}
Here, $[M]=  \prod_{n=1}^{N-1}[m]_n$, and $\delta_+$ and $\delta_-$ describe the terms generated by iteratative application of Eq. (6) that do not contain $B_N^-$.  $\delta_+$ and $\delta_-$ are determined by the values of $Q_n$. When calculating $[M]$ for a specific frequency $u_n$, to ensure numerical stability, we assume any layer where $u_nL_n > 3$ is thermally thick. In other words, for layers where $u_nL_n > 3$, we assume $B_n^+ \approx 0.$ Finally, applying the final heat-current boundary condition at the surface, ie. Eq. (\ref{boundary2}) with $n=1$, to eliminate $B_N^-$ yields:
\begin{equation}
\begin{bmatrix}
B_1^+\\
B_1^-
\end{bmatrix} = \frac{1}{M_{12}-M_{22}}(\frac{-Q_1}{\gamma_1}-\delta_++\delta_-)\begin{bmatrix}
 M_{12} \\ 
 M_{22}
\end{bmatrix} + \begin{bmatrix}
\delta_+ \\ 
\delta_-
\end{bmatrix}.
\label{surfTemp}
\end{equation}
\noindent
Here, $M_{ij}$ denotes the element in the $i^{th}$ row and $j^{th}$ column of matrix $[M]$. 

\noindent
\\
The temperature that determines the thermoreflectance signal depends on the experimental geometry. For experiments in a front/back geometry, like the one illustrated in FIG. 1, it is reasonable to assume thermoreflectance signals are proportional to the surface temperature of the multilayer. In this case, the Green's function that replaces Eq. (18) in Ref. \cite{cahill2004analysis} is:
\noindent
\begin{equation}
G(k,\omega)=B_1^++B_1^-.
\label{green}
\end{equation}
\noindent
We note that for the special case where all $Q_n$'s but $Q_1$ are zero, $\delta_+$ and $\delta_-$ are zero and Eq. (\ref{green}) is equivalent to Eq. (18) from Ref. \cite{cahill2004analysis}.

\noindent
\\
In some geometries, the thermoreflectance signal may be proportional to the temperature at the surface of a buried layer.  In this case, the temperature amplitudes for the appropriate buried layer (or layers) need to be computed from the surface temperature, Eq. (\ref{surfTemp}).  Matrix manipulation of Eq. (\ref{iteration}) yields:

\noindent
\begin{equation}
\begin{split}
\begin{bmatrix}
B_{n+1}^{+}\\ 
B_{n+1}^{-}
\end{bmatrix} =  \frac{1}{2\gamma_{n+1}}\begin{bmatrix}
(\gamma_{n+1}+\gamma_n)&  (\gamma_{n+1}-\gamma_n) \\
(\gamma_{n+1}-\gamma_n)& (\gamma_{n+1}+\gamma_n)
\end{bmatrix}\\ \times   
\begin{bmatrix}
e^{u_nL_n}&0 \\
0&e^{-u_nL_n}
\end{bmatrix}
\begin{bmatrix}
B_{n}^{+}\\ 
B_{n}^{-}
\end{bmatrix} \\-\frac{1}{2}  
\begin{bmatrix}
1&\gamma_{n+1}^{-1}\\
1& -\gamma_{n+1}^{-1}
\end{bmatrix}
\begin{bmatrix}
0\\
Q_{n+1}
\end{bmatrix}.
\label{f_iteration}
\end{split}
\end{equation}

\noindent
We now consider the general case.  In general, thermoreflectance signals are a weighted average of the temperature profile throughout the layered structure. 
\noindent
\begin{equation}
G(k,\omega)=\sum_{n=1}^{N} \int_{0}^{L_n} W_n(z)(B_n^+e^{u_nz}+B_n^-e^{-u_nz}) \,dx.
\label{green2}
\end{equation}.
\noindent \\
$W_n(z)$ is the weight function and depends on the geometry of the experiment, each layer's thickness and optical constants, and each layer's thermo-optic constants. Because $W_n(z)$ is not straightforward to analytically derive, we calculate it numerically. For a detailed description of how to calculate $W_n(z)$, see supplemental material of Ref. \cite{liu2021differentiating}.  Once $W_n(z)$ is known for each layer, we evaluate the integral in Eq. (\ref{green2}) numerically via Gauss quadrature:     
\noindent
\begin{equation}
    G(k,\omega) \approx  \sum_{n=1}^{N} \sum_{i=1}^{l}w_iW_n(z_i)(B_n^+e^{u_nz_i}+B_n^-e^{-u_nz_i}),
\end{equation}

\noindent
\\
where $w_i$ is the appropriate Gauss weight for node $i$ centered at depth $z_i$. 
\textcolor{black}{
In the example shown in FIG. 1, the weighting function is zero in all layers except the Al thermometer layer. If the thermometer layer is opaque, the weight function is zero in other layers. If the thermometer layer is not opaque, other layers can contribute to the signal. The extra complexity this causes is undesirable because the signal will depend on thermo-optic coefficients in addition to thermal properties.}

\noindent
\\
We conclude the description of the thermal model with three comments:

\noindent
\\
(i) While the heat equation in Eq. (\ref{heat_equation}) assumes a radially symmetric thermal conductivity, generalizing the above model for an anisotropic thermal conductivity tensor is straightforward.  Starting with a heat-equation for anisotropic materials will change  $u_n$, and will lead to slightly different matrices in Eq. (6) for relating temperature amplitudes, see the appendix in Ref. \cite{feser2014pump}.

\noindent
\\
(ii) Our goal for deriving the above solution is to interpret time-domain thermoreflectance experiments in a laser flash geometry, see FIG. \ref{schematic}.  However, the Green's functions derived above could be used to interpret data for a variety of other experiments, e.g. frequency-domain thermoreflectance experiments \cite{schmidt2009frequency}, steady-state thermoreflectance experiments \cite{braun2019steady}, beam-offset thermoreflectance experiments \cite{feser2012probing,feser2014pump}, time-domain thermoreflectance experiments with elliptical beams \cite{li2018anisotropic}, or Raman thermometery \cite{beechem2015invited,kargar2020phonon}. \\
\noindent
\\
\textcolor{black}{
(iii) Previously published thermal models for analyzing thermoreflectance assume the pump and probe beam interacts with the same plane \cite{cahill2004analysis,feser2014pump,feser2012probing,braun2019steady}, or the same layer \cite{ yang2016modeling}. Feser et al. allow the pump and probe beams to be separated radially, but not as a function of depth. The model described here allows pump and probe beams to be separated as a function of depth, which is necessary for analyzing TDTR experiments in a front/back geometry. Another advantage of the approach illustrated in FIG. 1 is it can be used to model arbitrary heating profiles. In multilayers with layers that are not optically thick, absorption vs. depth can be a complex function due to interference effects and not well described as a decaying exponential. Accurate modeling of the absorption profile is only necessary when studying samples whose thickness is comparable to the lengthscale heat is deposited over.}

\textcolor{black}{\section{Methods}
\noindent
To test thermal model predictions, we performed a series of front/front and front/back TDTR measurements on an Al/amorphous-Si/Al trilayer.  The 75nm Al/60nm amorphous-Si/75nm Al/sapphire stack was prepared via sputter deposition.  The Al films were DC magnetron sputter deposited.  The Si layer was radio-frequency sputter deposited.  The thickness of the sample’s layers was determined with a combination of picosecond acoustics and scanning electron microscopy.}

\noindent
\\
\textcolor{black}{
Time-domain thermoreflectance (TDTR) measurements were performed using a pump-probe system with a Mai Tai Ti:sapphire 780 nm laser with a repetition rate of 80 MHz. In the front/back setup, the pump beam impinges through the sapphire substrate and is absorbed by the bottom Al layer. The probe beam hits the top Al layer on the opposite side of the stack.
To overlap the pump and probe beam, we rely on a camera that is integrated into our setup \cite{gomez2020high}. The camera provides a microscope image of the sample surface, the reflected probe beam, and the transmitted pump beam. The position of the pump and probe beams on the sample is controlled through small adjustments to mirror angles or the positions of the objective lenses. For opaque samples in which no pump light transmits, we start our experiments with a semi-transparent sample, such as a 20nm Pt film on sapphire. We overlap the pump and probe beam on the semi-transparent sample. Then, we replace the semi-transparent sample with the opaque sample of interest. Finally, before collecting thermoreflectance data vs. time-delay, we adjust the pump beam position until the thermoreflectance signal is maximized. We performed experiments with modulation frequencies between 4 and 20 MHz. Additional details on the experimental apparatus can be found in Ref.\cite{gomez2020high}.
}

\noindent
\textcolor{black}{
The thermal model parameters used to simulate the thermal response of this a-Si sample are summarized in Table 1. The thermal boundary resistance between Al and the amorphous layer is small compared to the thermal resistance of the amorphous layer itself for 60nm Si. We set the interface conductance of the Al/Si interface to be 350 MW/(m-$^2$K), a typical value for clean Al/Si systems. The thermal model predictions are not sensitive to this value. 
The amorphous Si thermal conductivity was determined from standard front/front TDTR experiments. The Al thermal conductivity was determined using the Wiedemann Franz law and four-point probe measurements of electrical resistivity. The Al/sapphire interface conductance was determined by separating front/front measurements of a single Al film on sapphire. Heat capacities of all layers are set to literature values. To model the optical absorption of the pump beam, the heater layer is divided into 9 sub-layers. Metals typically have optical penetration depths of $\sim$ 10 nm. We found that division of the heating layer into sublayers with a thickness similar to the optical penetration depth is sufficient for convergence at time delays as short as 3 ps. Division of the heating layer into 9 sublayers increases the simulation time by only a factor of two.  }
\begin{table}[]
\textcolor{black}{
\caption{The thermal and optical parameters we used for to simulate thermal signals in FIG. 2-4. The same parameters were used for simulations for front/back and front/front \cite{cahill2004analysis} configurations. }
\renewcommand{\arraystretch}{1.5}
\begin{tabular}{r|c}
a-Si   thickness (nm)               & 60          \\\hline
a-Si thermal conductivity (W/(m-K))    & 0.5         \\\hline
a-Si heat capacity (J/(cm$^3$-K))    & 2.1      \\\hline
Al (heater) thickness (nm)        & 75          \\\hline
Al (thermometer) thickness   (nm) & 75 
\\\hline
Al heat capacity (J/(cm$^3$-K))    & 2.4      \\\hline
Al conductivity (W/(m-K))            & 120         \\\hline
sapphire thermal conductivity (W/(m-K))    & 36         \\\hline
Al/sapphire interface conductance (MW/(m$^2$-K))    & 170       \\\hline
sapphire heat capacity (J/(cm$^3$-K))    & 3.1     \\\hline
Laser spot size ($\mu$m)              & 6.2         \\\hline
Laser wavelength (nm)                   & 783         \\\hline
Modulation frequency (MHz)        & 4; 10.7; 20 \\\hline
Al refractive index &  1.4 + 5.7$i$ \\\hline
\end{tabular}}
\end{table}

\begin{figure*}[!ht]
  \centering
    \includegraphics[width=0.95\textwidth]{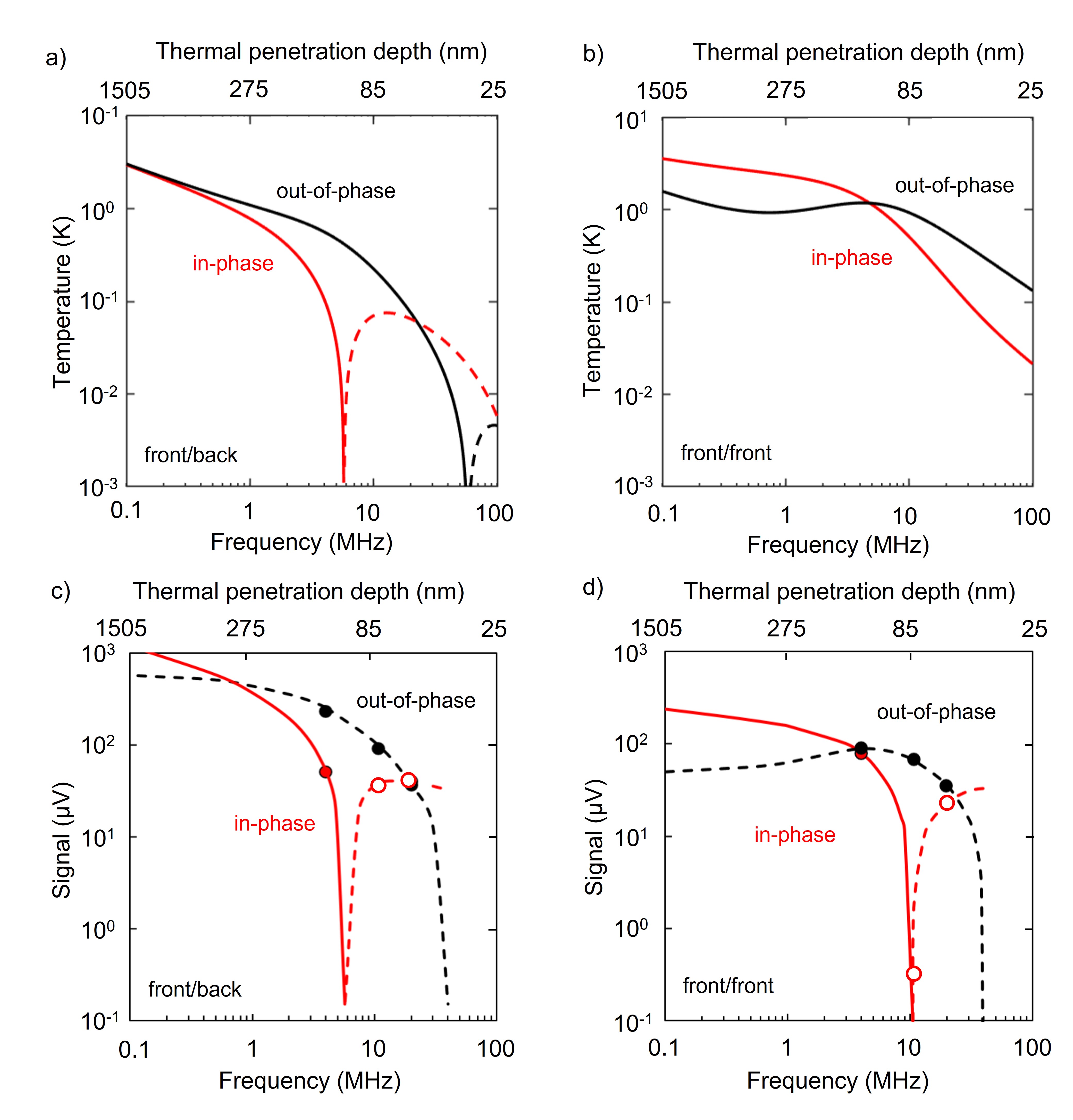}
    \caption{\textcolor{black}{ Temperature response of a one-dimensional a 75nm Al/60nm a-Si/75nm Al/sapphire multilayer as a function of heating frequency in a a) front/back configuration, and b) front/front configuration. In the front/back geometry, the sign of the in-phase and out-of-phase temperature oscillates with increasing frequency because the phase of the temperature response depends on $L_{Si}/d_p$. The dashed lines represent the negative of the original value. c) Experimental data (circle) and thermal model predictions (line) for the in-phase (red) and out-of-phase (black) TDTR signals of an Al/60nm Si/Al/Sapphire sample as a function of heating frequency in a c) front/back configuration, and d) front/front configuration. The laser spot size in these experiments is 6.2 $\mu$m. The thermal parameters we used for c) and d) are shown in Table 1. } }
     \label{frequency_fb_ff}
\end{figure*}

\begin{figure}[!ht]
  \centering
    \includegraphics[width=0.45\textwidth]{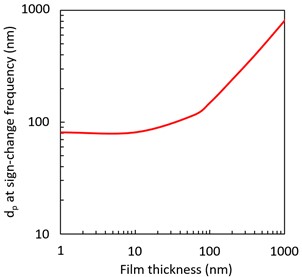}
    \caption{\textcolor{black}{Relationship between the sign-change frequency in the in-phase TDTR data and the thickness of the a-Si layer. The y-axis shows the thermal penetration depth of a thermal wave in a-Si, $d_p$, at the heating frequency where the sign change in the in-phase TDTR signal occurs. These calculations are for a 75nm Al/a-Si/75nm Al/sapphire multilayer. For a 60 nm thick a-Si, the sign change occurs at $\sim$ 6 MHz (FIG. 2c), which corresponds to $d_p$ = 112 nm. For thick layers, $d_p$ at the sign change frequency is nearly equal to the film thickness. }}
     \label{d_p_thickness}
\end{figure}

\begin{figure*}[!ht]
  \centering
    \includegraphics[width=1.0\textwidth]{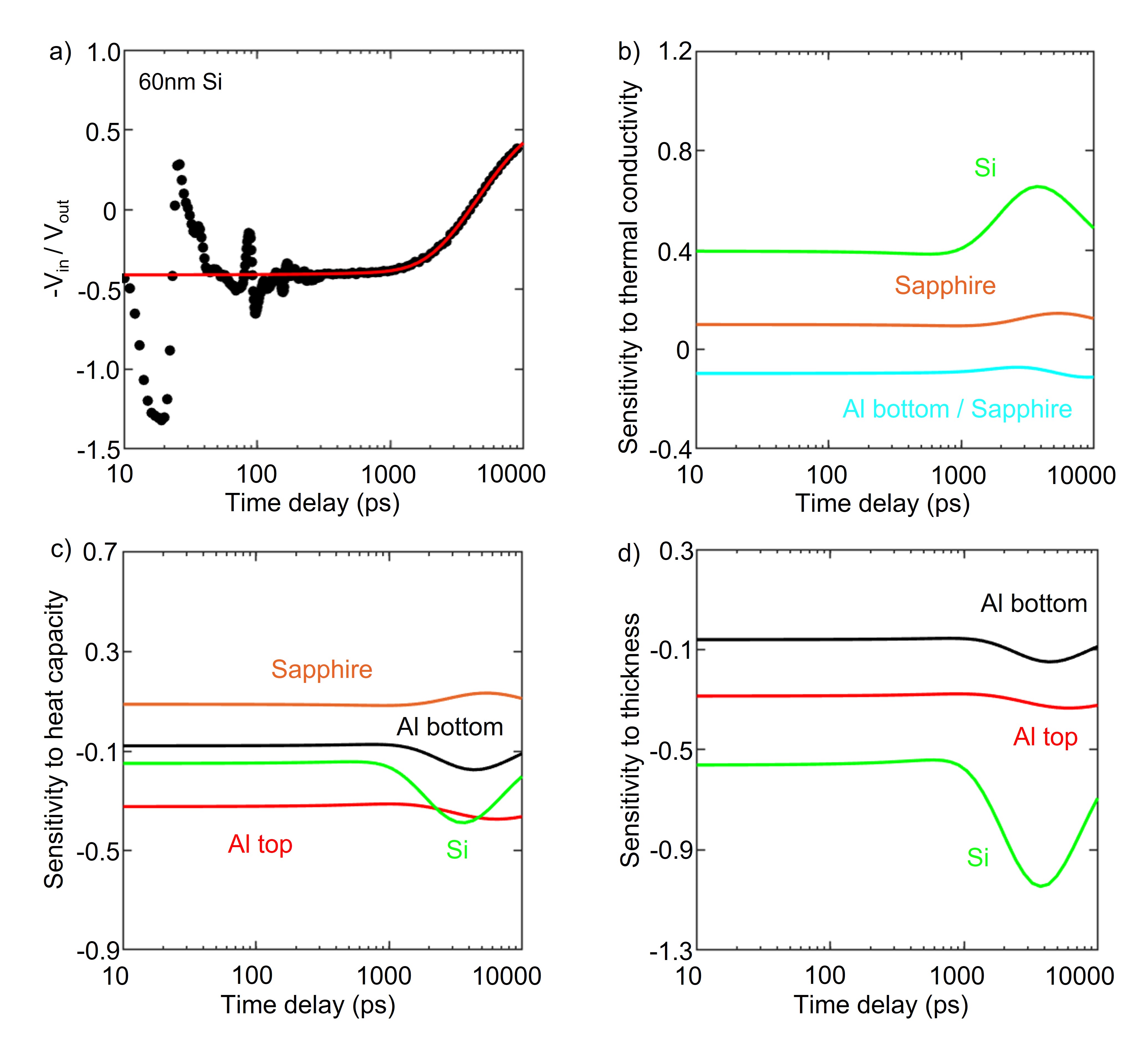}
    \caption{\textcolor{black}{a) Experimental data (circles) and thermal model predictions (lines) for front/back TDTR of a 75nm Al/60nm a-Si/75nm Al/sapphire multilayer. The thermal parameters we used are shown in Table 1.   b) The sensitivity of experimental signals of the 60nm Si is shown in a) to the thermal conductivity and interface conductance of the a-Si and sapphire. c) The sensitivity of experimental signals to the heat capacity of sapphire, top and bottom Al layers, and a-Si layer. d) The sensitivity of experimental signals to the thickness. The maxima in sensitivity near $\approx$ 4 ns for many of the curves is because these parameters help determine the thermal relaxation time of the Al/a-Si/Al tri-layer. Additionally, the sensitivity to $\Lambda$ is opposite in sign as the sensitivity to $C$ and $L$. Intuitively, the thermal diffusion across a layer of thickness $L$ scales with $L^2C/\Lambda$. We expect the sign of the sensitivity of $-V_{in}(t_d)$ to be negative for $L$ and $C$ and positive for $\Lambda$. } }
     \label{fb_sensitivity}
\end{figure*}

\begin{figure*}[!ht]
  \centering
   \includegraphics[width=0.9\textwidth]{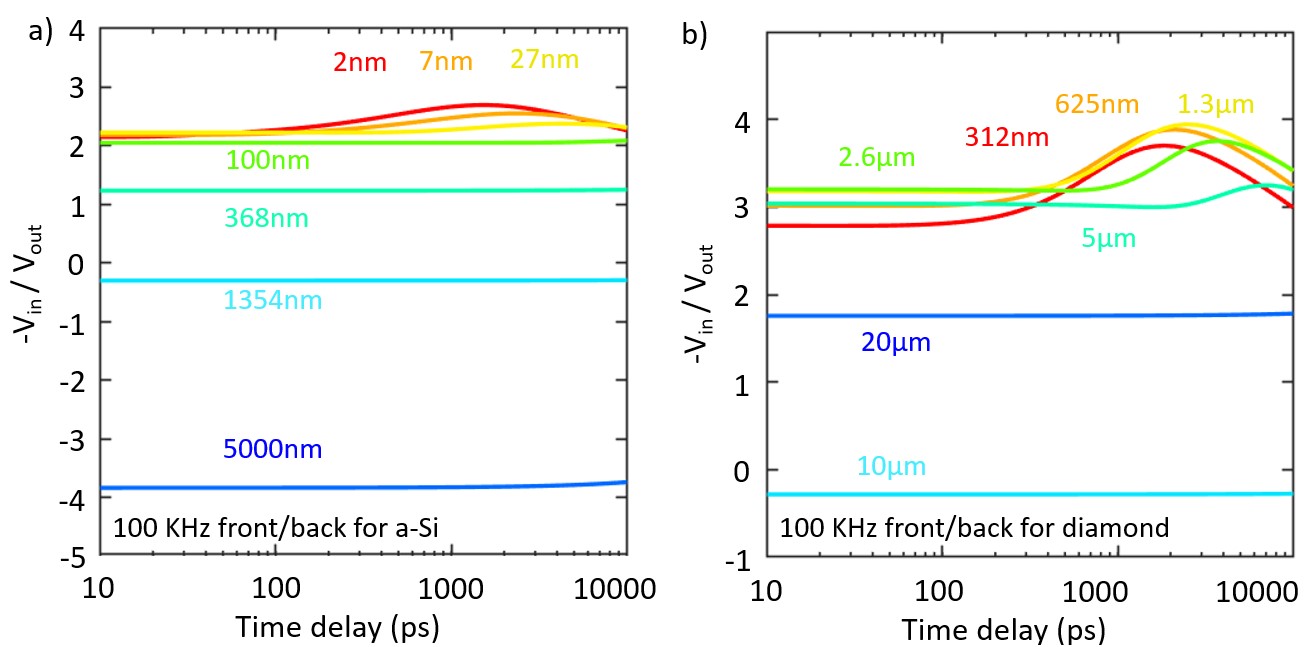}
   \caption{\textcolor{black}{TDTR signal as a function of film thickness for a) a-Si and b) diamond transport layers. For films that have a thermal response time shorter than 10 ns, a laser-flash-like response can be observed. For films too thick to have a laser-flash-like signal vs. time-delay, there is a time-delay independent signal from pulse accumulation effects. The thermal conductivity for the diamond is set to be 2000 W/(m-K). The interface conductance of Al/diamond is set to be the same as Al/a-Si, 350 MW/(m$^2$-K), for comparison purposes.}}
    \label{thickness_sensitivity_ff_fb}
\end{figure*}
\section{Results}
\noindent
We now use Eq. (\ref{green}) to analyze the thermal response of an amorphous Si thin-film to laser heating. We begin our analysis by considering the frequency domain thermal response of a 1D Al/amorphous Si/Al trilayer, see FIG. 2a and 2b. Then, we consider frequency-dependent TDTR data (FIG 2c-d, and FIG. 3). Finally, we compare our model's predictions for TDTR data vs. time-delay to experimental data (FIG. 4). 

\noindent
\\
FIG 2a and 2b compare thermal model predictions for the frequency-domain temperature response of the a-Si sample to harmonic heating in a front/back vs. front/front geometry. The frequency-domain response of the same sample in these two different experimental configurations is strikingly different. The temperature in a front/front geometry decreases slowly with increasing heating frequency (FIG. \ref{frequency_fb_ff} b).  In a front/back geometry, the temperature decreases more rapidly with frequency (FIG. \ref{frequency_fb_ff} a). In the front/front geometry (FIG. \ref{frequency_fb_ff} b), the sign of the in-phase and out-of-phase temperature response is always the same and similar in magnitude. There is only a small amount of structure near a frequency of 5 MHz. This structure in the frequency-domain thermal response is related to the thermal penetration depth of a-Si. Alternatively, in the front/back geometry, both the in-phase and out-of-phase temperature response change sign periodically with increasing frequency.

\noindent 
\\
Why does \textcolor{black}{the magnitude and the sign} of the in-phase and out-of-phase temperature depend strongly on frequency in a front/back experiment, \textcolor{black}{but} not in a front/front experiment? To answer this question, we consider the solution to the heat diffusion equation for a semi-infinite solid in 1D in response to harmonic surface heating:
\begin{equation}
  \Delta T(z,\omega) = Q_1\frac{d_p}{\Lambda}e^{-z/d_p}(1-i)e^{-iz/d_p},   
\end{equation}
%
in which $d_p$ is the thermal penetration depth and $z$ is the depth from the sample surface. 
In a front/front TDTR experiment, $z$ = 0.  Then, the in-phase and out-of-phase will not change sign, and the frequency dependence comes from only $d_p$, which is proportional to $\omega^{-1/2}$.  Alternatively, if $z = L$, the frequency dependence of both the amplitude and phase change is a function of frequency.  The amplitude of the temperature response now includes a $e^{(-L/d_p)}$ factor that causes the temperature response to decay exponentially with increasing frequency.  Similarly, the $e^{(-iz/d_p)}$ factor causes both the real (in-phase) and imaginary (out-of-phase) thermal response to change signs with increasing frequency.  While the tri-layer we report on in FIG. \ref{frequency_fb_ff} is more complicated than the semi-infinite solid, the behavior is qualitatively similar to what we describe above.

\noindent
\\
\textcolor{black}{We now turn our attention to TDTR signals as a function of modulation frequency. In FIG. 2c and 2d, we show predictions for the TDTR signal at a pump/probe delay time of -100 ps. The signals for time-domain thermoreflectance were calculated using Eq. (9) to generate the frequency domain response of the sample. Then, the TDTR signal was constructed from a linear combination of the appropriate frequencies, as outlined in Ref. \cite{cahill2004analysis}. We include experimental data collected at 4, 10.7, and 20 MHz for comparison with the thermal model predictions. The experimental data were scaled by a frequency-independent constant to match the magnitude of the theoretical predictions at 20 MHz.}
\noindent\\
The dependence of the TDTR signals on modulation frequency is similar to the trend we observe in FIG. 2a and 2b for the frequency domain. The in-phase signal in FIG. 2c changes sign at nearly the same frequency as the frequency-domain temperature changes sign in FIG. 2a. Additionally, the magnitude of the thermal response decays with frequency more quickly in a front/back than in a front/front experiment. \\

\noindent\\ \textcolor{black}{ While many of the trends in FIG. 2a and b are similar to c and d, there are several notable differences. These are related to pulse accumulation effects in the TDTR signal. For example, there is a sign change in the front/front in-phase TDTR signal near 10 MHz (FIG. 2d) that is unrelated to the thermal properties of the sample and has no parallel in the frequency domain response (FIG. 2b). Pulse accumulation leads to a sign change in the TDTR in-phase signal at negative delay times whenever the modulation frequency is 1/8$^{th}$ the rep-rate of the laser\cite{cahill2004analysis}. Our laser rep frequency is 80 MHz. Therefore, the sign change happens at about 10 MHz. }

\noindent\\ 
We emphasize that the sign change in the front/back in-phase TDTR signal near 5 MHz (FIG. 2c) is a thermal penetration depth effect, unlike the sign change in the front/front geometry (Fig. 2b). To verify this, we used our thermal model to calculate the sign-change frequency as a function of a-Si thickness. We show the results of these calculations in FIG. 3. In FIG. 3, the cross-over frequency is plotted in units of length using the formula $d_p=\sqrt{\alpha /(\pi f_{mod})}$. As expected, when the a-Si is thin, the cross-over frequency does not depend on the properties of the a-Si layer because the a-Si layer is not an important thermal resistance. Once the film is appreciably thick, e.g., greater than 50 nm, the cross-over frequency provides a measure of the thermal penetration depth of the a-Si layer.

\noindent
\\
We now turn our attention to how TDTR signals evolve as a function of pump/probe delay time. In FIG. \ref{fb_sensitivity} a), we show front/back TDTR data and thermal model predictions for the Al/a-Si/Al sample as a function of time-delay. As expected for an experiment conducted in a laser flash geometry, we observe a constant signal at delay times less than the time-scale it takes for heat to diffuse across the Al/a-Si/Al tri-layer. Then, the signal increases as heat diffuse into the top Al thermometer layer. The oscillations in the TDTR data on sub-nanosecond time-scales (FIG.\ref{fb_sensitivity} a) are picosecond acoustics signal. 

\noindent
\\
To better understand how the thermal properties determine experimental signals in a front/back experiment, we perform a sensitivity analysis. Normally, sensitivity in a TDTR experiment is defined as the percent change in ratio due to the percent change in a certain model parameter $\beta$. That definition is problematic for front/back experiments because the ratio often crosses through zero, see FIG. \ref{fb_sensitivity}. Therefore, we define the sensitivity function S$_\beta (t_d)$ as the absolute change in ratio due to a small fractional change in $\beta$. \\
\begin{equation}
    S_\beta (t_d) =\frac{\Delta(-V_{in}(t_d)/V_{out}(t_d))}{\Delta\beta/\beta}, 
\end{equation}

\noindent
\\
\textcolor{black}{FIG. \ref{fb_sensitivity} b)-d) shows sensitivity curves for a 60 nm Si sample with a 75 nm Al heater and a 75 nm thermometer. } \\
Many of the sensitivity curves in FIG. \ref{fb_sensitivity} have a similar shape for time delays between 1 and 10 ns. This similarity in shape is because the front/back TDTR experiment measures the time-scale it takes for heat to diffuse from the heater to the thermometer layer. 
\textcolor{black}{
The maximum sensitivity to the thermal conductivity and thickness of the a-Si is at $\sim$ 4 ns. This means the thermal relaxation of the trilayer is $\sim$ 4 ns. The thermal relaxation time in a normal laser flash experiment is $\tau \approx 0.14\times (L/\Lambda)(LC) = 0.14\times L^2/\alpha$.  Here, $L/\Lambda$ is the thermal resistance of the layer, $LC$ is the heat capacity per unit area of the layer, and  $\alpha$ is the thermal diffusivity of the material. In our experiment, the thermal relaxation time is influenced by the heat capacity of the metal thermometer layer. In theory, it also depends on the thermal resistance of the metal layers and the boundary resistance between the metal and a-Si. But in practice, these thermal resistances are negligibly small. Therefore, a crude estimate for the thermal relaxation time of the trilayer is  $\tau \approx 0.14\times (L_{a-Si}⁄\Lambda_{a-Si})(L_{Al}C_{Al}+L_{a-Si}C_{a-Si}) \approx 5$ ns, which is similar to the value for which we observe a maximum in the sensitivity, see FIG. 3.}



\noindent
\\
In addition to the laser-flash like transport that causes the ``hump" in the sensitivity curves near $\approx4$ ns, front/back experimental signals include effects from pulse accumulation \cite{schmidt2008pulse}. The temperature rise due to pulse accumulation leads to additional sensitivity to the stack's many thermal properties that is time-delay independent. 

\noindent
\\
\textcolor{black}{
We now discuss the minimum and maximum sample thicknesses a front/back TDTR experiment can measure thermal properties of. The minimum thickness is set by the length scales the pump beam deposits heat, and the probe beam measures temperature. Careful experiment design can allow samples to be as thin as a few nanometers. For example, Ref. \cite{jang2020thermal} measure thermal transport across tunnel junctions a few atomic layers thick in a front/back geometry like illustrated in FIG. 1.}

\noindent
\\
\textcolor{black}{
The maximum thickness in a front/back experiment depends on the maximum delay time between pump and probe beams and the pump modulation frequency. As discussed above, there are two parts of the signal in a front/back TDTR experiment. There is a laser-flash-like signal that is similar to the thermal response that would occur from a single pulse. There is also a signal from accumulated pulses \cite{schmidt2008pulse}. \\
\noindent
\\
Our experiments have a maximum time delay between pump and probe pulses of $\sim$ 10 ns. Thermal diffusivity of solid materials typically falls between 10$^{-6}$ m$^2$/s (amorphous materials) and 10$^{-3}$ m$^2$/s (diamond). Therefore, depending on $\alpha$, the maximum thickness for which a laser flash signal is observable will be between 100 nm and 10 $\mu$m.  To verify this, we use our thermal model to predict the thermal response of a-Si layers and diamond layers with varied film thickness, see FIG 5. The thermal parameters we use for the a-Si simulations in FIG. 5a are shown in Table 1.  The only change is the film thickness and pump modulation frequency. Since we are interested in how thick the sample can be, we set the modulation frequency to be 100 kHz so the thermal penetration depth is as long as possible. (In most TDTR systems that use Ti:sapphire oscillators, laser noise makes experiments with modulation frequencies below 100 kHz challenging.) The stack geometry and experimental parameters we use for the diamond calculation in FIG. 5b are the same as for the a-Si layer, but with a transport layer thermal diffusivity set equal to 1.1$\times$10$^{-3}$ m$^2$/s.  And, the Al/diamond boundary conductance is set to 350 MW/(m$^2$-K).  As expected, the laser-like flash signal goes away for the a-Si or diamond layers when the film thickness approaches $\sim$100 nm and $\sim$10 $\mu$m.  
}

\noindent
\\
\textcolor{black}{
For films too thick to have a laser-flash-like signal vs. time-delay, there is still a signal from pulse accumulation effects \cite{schmidt2008pulse}. In this case, the signal becomes independent of time-delay, see results for thick layers in Fig. 5.  The signal from pulse accumulation describes thermal responses on timescales as long as $1/f_{mod}$. Therefore, the maximum thickness is governed by the thermal penetration depth  $d_p=\sqrt{\alpha /(\pi f_{mod})}$.  Samples much greater than this thickness will not have observable thermoreflectance signals. 
}


\noindent
\section{Conclusion}
\noindent
In summary, we presented a thermal model for the analysis of TDTR experiments which allows the presence of non-homogeneous optical heating. We performed front/back experiments on Al/a-Si/Al/sapphire to test our model. In the frequency domain, the magnitude of the in-phase and out-of-phase temperature of a front/back measurement changes dramatically and oscillates between positive and negative as the heating frequency increases. The period of oscillation is governed by the thermal penetration depth. Therefore, frequency-dependent experiments in a front/back geometry provide a direct measure of the thermal penetration depth in the multilayer. Finally, we examined the sensitivity of the front/back measurements experiments in a front/back geometry by measuring the thermal relaxation time of the tri-layer. 
\section*{Acknowledgement}
\noindent
This research was supported as part of ULTRA, an Energy Frontier Research Center funded by the U.S. Department of Energy (DOE), Office of Science, Basic Energy Sciences (BES), under Award \# DE-SC0021230.

\bibliography{apssamp}
\end{document}